\documentclass[conference]{IEEEtran}
\IEEEoverridecommandlockouts
\usepackage{cite}
\usepackage{amsmath,amssymb,amsfonts}
\usepackage{algorithmic}
\usepackage{graphicx}
\usepackage{textcomp}
\usepackage{xcolor}
\usepackage{float}
\usepackage{svg}
\usepackage{soul}
\usepackage{booktabs}
\usepackage{todonotes}
\usepackage[hidelinks]{hyperref}
\usepackage{array}

\def\BibTeX{{\rm B\kern-.05em{\sc i\kern-.025em b}\kern-.08em
    T\kern-.1667em\lower.7ex\hbox{E}\kern-.125emX}}
\renewcommand\baselinestretch{0.96}

\begin{document}


\title{FAV-NSS: An HIL Framework for Accelerating Validation of Automotive Network Security Strategies}
\author{
Changhong Li, 
Shashwat Khandelwal,
Shreejith Shanker\\
Reconfigurable Computing Systems Lab, Electronic \& Electrical Engineering\\
Trinity College Dublin, Ireland\\
Email:\{lic9, khandels, shankers\}@tcd.ie
}

\maketitle
\begin{abstract}
Complex electronic control unit (ECU) architectures, software models and in-vehicle networks are consistently improving safety and comfort functions in modern vehicles. 
However, the extended functionality and increased connectivity introduce new security risks and vulnerabilities that can be exploited on legacy automotive networks such as the controller area network (CAN).
With the rising complexity of vehicular systems and attack vectors, the need for a flexible hardware-in-the-loop (HIL) test fixture that can inject attacks and validate the performance of countermeasures in near-real-world conditions in real time is vital. 
This paper presents an FPGA-based HIL framework tailored towards validating network security approaches (IDS, IPS) and smart integration strategies of such capabilities for an automotive CAN bus.
FAV-NSS replicates an actual vehicular system environment with functional ECUs and network infrastructure on an FPGA, allowing functional validation of IDS/IPS algorithms, accelerator designs and integration schemes (software task on ECU, dedicated accelerator).
Software APIs on the attached host machine control and configure ECUs, automate test case execution and log signals from the ECUs and the `virtual' CAN bus during runtime.
To show the efficacy of FAV-NSS, we evaluate an IDS accelerator integration problem, both as a traditional coupled accelerator (to the ECU), and secondly close to the CAN controller (mimicking an extended CAN controller).
We show that the latter strategy can be fully validated by our framework, which would otherwise require integration of specialised CAN modules into otherwise standard HIL fixtures with ability to instrument internal signals for characterising timing performance. 
The tests demonstrate a promising latency reduction of 6.3$\times$ when compared to the traditional coupled accelerator.
Our case study demonstrates the potential of FAV-NSS for accelerating the optimisation, integration and verification of smart ECUs and communication controllers in current and future vehicular systems.

\end{abstract}

\section{Introduction}\label{introduction}
Recent years have seen rapid adoption of intelligent systems in production vehicles that improve the safety, reliability, and comfort of users~\cite{nidamanuri2021progressive}.
The advancements are enabled by a number of networked electronic control units (ECUs) that run software tasks to observe, monitor, and control different sensors and actuators in real-time. 
Typically, over 100 ECUs are present in modern cars~\cite{ecunum} that exchange information over robust communication protocols such as local interconnect network (LIN), FlexRay, controller area network (CAN), and Automotive Ethernet.
Despite numerous security issues, CAN continues to be the most widely used network protocol for critical and non-critical functions in modern vehicles. 

Development of automotive functions are typically done in silos and integrated into test fixtures for validating their functionality prior to deployment in real systems. 
%
%
%
Hardware-in-loop (HIL) simulation/emulation has been used for real-time testing of embedded systems, especially in the automotive and aerospace areas~\cite{shreejith2013accelerating,rao2022modeling}.
Multiple research has shown that field programmable gate arrays (FPGAs) are optimal target platforms for a flexible HIL setup.
FPGA-based HIL systems allow rapid exploration and validation of design choices through real-time emulation and reprogramming, rather than requiring expensive rewiring and re-validation with fixed platform-based HIL models.
HIL setups are also adopted in automotive systems development, primarily for functional-level and system-level verification of new features.
With increasing connectivity in automotive systems, system and network security verifications are increasingly gaining attention in the automotive domain. 
Unlike functional and system-level tests, an HIL system for in-vehicle safety research should be capable of rapidly exploring design solutions and verifying the constructed safety methods for myriad network conditions and ECU functions.
It should provide means for injecting (new) attack features and real-time recording and inspection, allowing end-to-end verification of security countermeasures and quantify their impacts on key ECU functions.
Techniques such as quantised neural networks based IDS~\cite{khandelwal2023exploring} provide unique integration strategies for security measures, moving them closer to the network controller, as opposed to traditional ECU-coupled and software IDS solutions. 
Validating the integration of these functions (in software/hardware) would require extensive modifications in a fixed HIL environment, and hence, impose restrictions on investigating novel security schemes, integration strategies and their validation in vehicular networks.

This paper presents a HIL framework for testing and validating network security schemes (such as IDSs \& IPSs) for CAN-based vehicular systems, with an FPGA as the heart of the framework. 
Our CAN testbed uses an Artix-7 FPGA to emulate multiple ECUs that are interconnected with a `virtual' CAN bus on the logic, with the ability to expand to multi-FPGA setups for scalability. 
The testbed can be controlled and configured through both a GUI-based or an API-based interface, and supports injection of attack vectors (including preconfigured ones such as DoS), replay of captured sequences (for replicating datasets), programming ECUs, and real-time acquisition and monitoring and analysis of CAN bus data and ECU status signals. %
To demonstrate the potential of the platform, we investigate two case studies which present different IDS integration pathways -- first, as a coupled accelerator to the ECU and second, as an integration closer to the CAN controller of the ECU. 
In both cases, we use an identical quantised neural network as the IDS to show that the platform uniquely provides the ability to evaluate the difference in integration strategies, in terms of detection performance and latency, in addition to the standard functional verification and end-to-end validation. 
The major contributions of this paper are as follows:

\begin{itemize}
    \item An FPGA-based hardware-in-loop testbed with configurable parameters for accelerating validation and testing of CAN IDS/IPS at both the ECU and networked function levels, while capturing ECU/function-level performance impacts due to IDS/IPS.
    \item Case study demonstrates the adaptability of the HIL platform to two different integration strategies for intrusion detection systems: the conventional coupled accelerator and the novel IDS-enabled CAN controller. We show that either integration can be tested and qualified against a range of attack vectors that can be injected under software control (GUI/API) on FAV-NSS.
    \item Additionally, the timing characterisation of the integration strategy on FAV-NSS shows that IDS integrated close to a CAN controller reduces the detection latency by 6.3$\times$, allowing line-rate detection on most CAN networks.
\end{itemize}

The rest of this paper is organized as follows. Section~\ref{background} provides background and related research works on CAN security and mitigation schemes as well as on hardware-in-loop testbed and components; Section~\ref{architecture} describes the system architecture and implementation of the hardware-in-the-loop testbed, the software framework and the IDS model used for the case study. Section~\ref{casestudy} outlines the different integration methods for the IDS, with section~\ref{results} capturing the observed results from testing across different configured and programmable attacks, as well as the latency gains enabled by the different integration strategies. Finally, we conclude the paper in section~\ref{conclusion}.

\section{Background and related works}\label{background}

\subsection{CAN security threats \& mitigations}

\begin{figure}[t]
  \begin{center}  \includegraphics[width=1\linewidth,trim={0 0.6cm 0 0.9cm}, clip]{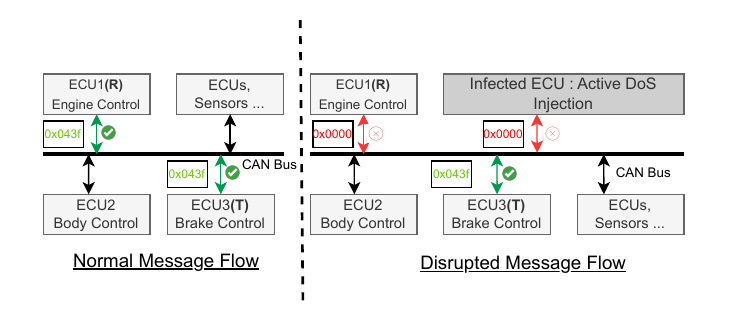}
  \end{center}
  \caption{Figure shows a compromised ECU using an active DoS attack to block critical message communication. In the figure, the engine control ECU fails to receive a message from the body control module due to the active DoS injections on the bus.}
  \label{attack} 
\end{figure}

CAN provides numerous advantages over competing protocols such as resistance to electromagnetic interference (due to twisted wire transmission), integrated arbitration and priorities, and its linear broadcast bus topology leading to lower overall cabling weight and cost.
However, due to increased connectivity to the external systems, some of these built-in properties can be exploited for network intrusion attacks, affecting the safety and reliability of the systems interconnected by CAN~\cite{bozdal2020evaluation}.
Vulnerabilities in the CAN protocol have been thoroughly evaluated in the research literature~\cite{9974508,song2020vehicle,9443234}.
The broadcast nature allows all ECUs to receive communication from any known/unknown/infected ECU on the network, passively sniff data on the network to identify/record unique information or reuse the sensor/actuator commands for replay attacks. 
The built-in arbitration scheme could be exploited for targeted attacks or simple flooding and DoS attacks, completely crippling the network. 
In~\cite{wolf2004security}, the authors were the first to formulate a theoretical risk level corresponding to the level of attacks that could be injected into vehicles. 
Further in~\cite{cosimi2022analysis,hoppe2008security}, the authors showed that it was possible to inject information frames into the vehicle control bus to bypass \textit{driver} and \textit{system} control to maliciously take over critical functions in the vehicle.
In~\cite{palanca2017stealth}, authors proposed a targeted DoS attack method on the CAN bus, which can isolate specific ECUs on the network and prevent them from communicating.
While many attack vectors could exploit these vulnerabilities in CAN, the most widely reported ones in the literature are \textit{DoS}, \textit{fuzzing} and \textit{targeted spoofing} attacks~\cite{song2020vehicle}.
DoS attacks can be launched on the CAN bus through \textit{bus flooding}, exploiting the \textit{error propagation} feature or by preventing the \textit{sleep-wake} sequence causing the ECUs to stay in sleep mode.
Bus flooding attack, shown in Fig.~\ref{attack}, is the most common DoS attack where the malicious ECU floods the bus with high-priority messages, preventing all other ECUs from using the bus.
Error propagation attacks are more sophisticated, where the error handling scheme in CAN is exploited by intentionally causing bit stuffing errors, which triggers listening ECUs to send error frames, thus stifling the bus. 
Fuzzing attacks use random payloads to disrupt normal functionality or to identify vulnerabilities in a specific ECU by observing their response to the payload. 
Targeted spoofing attacks use specific payloads aimed at one or more components, forcing them to perform incorrect operations or generate incorrect responses when polled. 

To address these issues, multiple IDS approaches have been proposed in the research literature.
Although many rule-based and signature-based IDSs were initially suggested, deep neural network (DNN)-based IDS solutions have increasingly gained traction, as they have shown to be more accurate than rule- and signature-based IDS solutions~\cite{khandelwal2022deep,9443234,wang2018entropy,shahriar2022canshield}.
Optimisations such as quantisation and pruning are often employed to make these DNN-based IDSs deployable in resource-constrained in-vehicle environments~\cite{khandelwal2023exploring}. 


\subsection{CAN testbeds}
CAN testbeds have been proposed and developed to model, test and validate CAN-based ECUs. 
In~\cite{wasicek2014virtual}, the authors proposed a virtual CAN overlay that abstracts the communication interface of the Multiprocessor System-on-Chip (MPSoC) and provides programmers with an application programmer interface (API) for interacting with the CAN network.
The work in~\cite{obermaisser2005ordering} presents a scheme to establish identical time base and message order in a virtual CAN network as the real (physical) one.
A lightweight CAN virtualisation is proposed in~\cite{lee2019virtualization} for virtual controllers to improve the functional and Quality of Service(QoS) issues with prior works by reducing the virtualisation overhead by 20\%.
A novel device-level virtualisation is proposed in~\cite{kim2011fieldbus} allowing CAN to be deployed in Integrated Modular Avionics (IMA) architecture. 
However, most of the virtualisation schemes emulate the CAN layer and are not compatible with external hardware CAN instances or sensors. 
This limited the scalability and adaptability of these simulation-based schemes for real-time multi-ECU test setups. 
Similar to network virtualisation, ECU virtualisation has also been explored for developing and optimising ECU architectures for performance and safety~\cite{niimi2012virtual}, functional validation of automotive software applications~\cite{mauss2014chip, bidkar2021virtual} among others. 
For large-scale testing, the work in~\cite{urbina2015multi} showed the use of virtual ECUs as abstractions integrated into a network-on-chip environment that is compliant with AUTOSAR specifications. 
Our work takes inspiration from these approaches but uses a RISC soft-core processor-based system as the ECU core with CAN controller logic as a memory-mapped IP for each ECU; ECUs are subsequently interconnected through a virtual CAN bus to model the systems as closely as possible to the real world. 

\subsection{HIL Test Setups}
Hardware-in-the-loop simulators/emulators are widely used for embedded system performance evaluation systems to accelerate the testing and validation of the performance of the system during development or revision phases in an environment that closely matches real-world settings.
Early reported use of HIL simulators was in the development and testing phases of fly-by-wire systems, flight simulation~\cite{dillard1998real} and missile guidance systems~\cite{pace1998effectiveness}, and for subsystem level verification of components of spacecraft systems~\cite{leitner1996space}.
Several R\&D efforts in the automotive domain subsequently adopted HIL simulations to verify ECU architecture, functionality and for safety testing~\cite{hanselmann1996hardware, puschmann2021safe}.
Extensive software models have been used to recreate system dynamics (e.g., of engines) when validating ECU functions in an HIL setup~\cite{rabbath2002simulating}.
Such frameworks often tend to be specialised (and thus relatively inflexible), requiring different setups to validate specific capabilities -- for e.g., setups for protocol-level development/changes are inflexible to be adapted for end-to-end functional validation of ECU architecture/software components. 
Additionally, capturing low-level details leads to highly complex software models that force the tests to be run at a slower speed than real-world deployment conditions. 
Alternatively, some models abstract away low-level details to improve speed of simulation/emulation, and in both cases lead to additional development time for functional, safety and integration tests at the system level. 
Field programmable gate arrays (FPGAs) have increasingly become a key component of such systems, allowing much simpler reconfiguration of the setup for different testing/development scenarios while also offering real-time performance as the dedicated hardware-based setups~\cite{bullock2004hardware,shreejith2013accelerating,mihalivc2022hardware}.
Our work builds on this approach, with a focus on enabling rapid prototyping and validation of (hardware-) accelerated network security approaches (IDS, IPS, smart controllers, integration strategies and others) for in-vehicle networks. 

\color{black}
\section{System architecture}\label{architecture}
\begin{figure}[t]
  \centering
\includegraphics[width=\columnwidth]{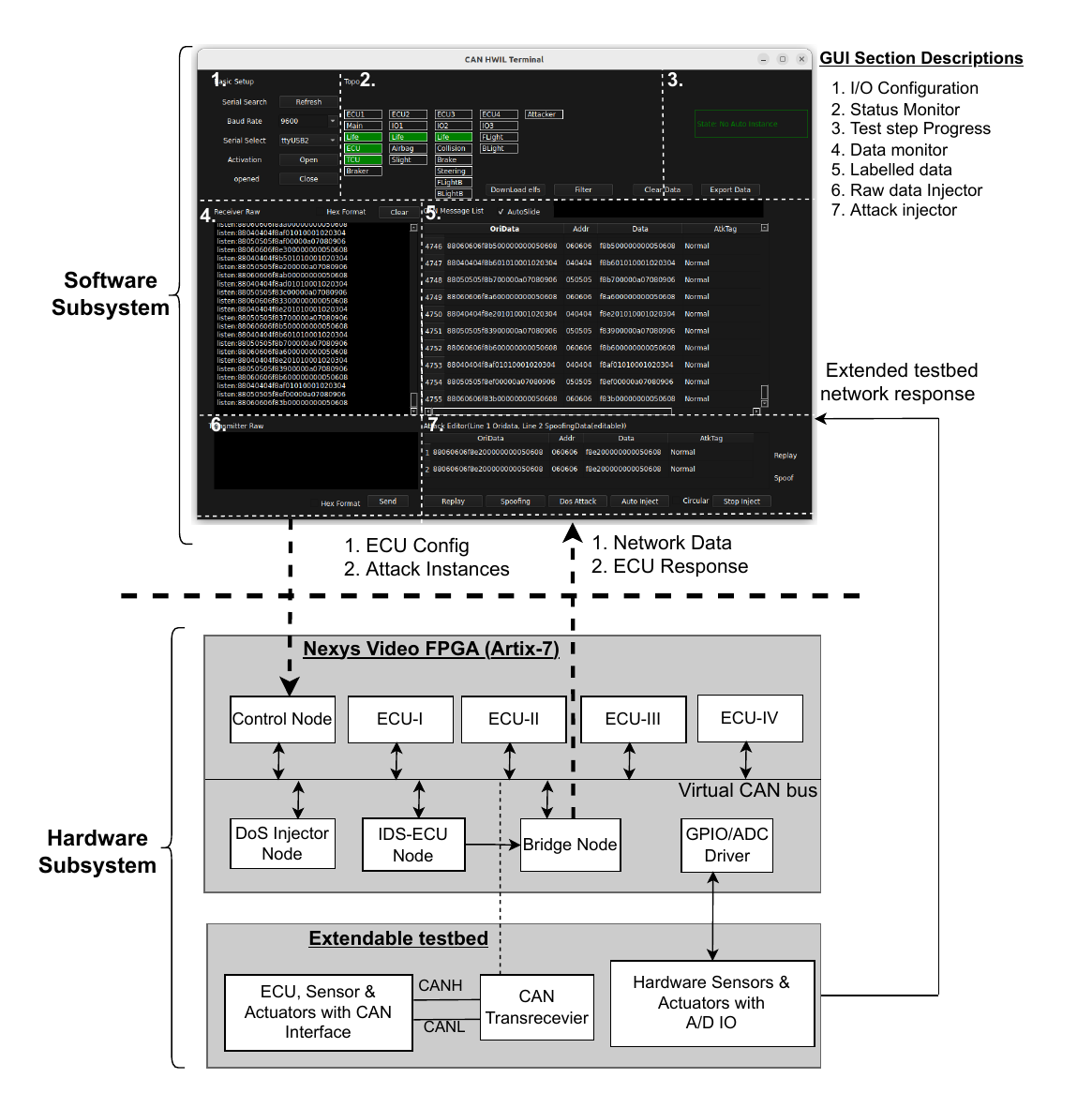}
  \caption{Overview of the architecture of the hardware-in-the-loop simulator showing how the different components interact with each other.}
  \label{fig3}
\end{figure}

In this section, we present the system architecture of the HIL framework and the CAN testbed. 
The high-level overview of the framework is shown in Fig.~\ref{fig3}.
The framework is comprised of a software GUI front-end which interacts with the CAN testbed that is implemented on an FPGA.
The software can access pre-prepared
test profiles, ECU application object code and relevant test cases from a pool of resources which can be used to set up and automate the test cases. 
The interface provides functionality to download the test configuration to the testbed, download target code to the ECUs, execute the test cases (attacks), perform real-time monitoring of the test and generate data dumps for detailed analysis. 
For this development, we have used AMD's Artix-7 XC7A200T FPGA on the Digilent Nexys Video kit.
The large FPGA can accommodate numerous soft-core processors with their own peripherals and CAN network interface (mimicking full-fledged ECUs) as well as dedicated blocks for injecting attacks, controlling the test setups and for real-time observation/monitoring. 
Each ECU runs its own application(s) independently and uses the CAN interface to communicate with other ECUs on the network.  
In addition to the software inputs, the dedicated I/O on the Nexys board is also mapped as control and status inputs to each of the ECUs.
%
We discuss the various components of the testbed in detail in the next sections.

\subsection{Software Subsystem}
The hardware testbed is controlled and configured primarily through a graphical interface we developed using PyQt. 
Each action in the GUI, partitioned into one of the 7 sections on the GUI as seen in Fig.~\ref{fig3}, uses a set of APIs defined in Table~\ref {tab:interface_function} to interface with the hardware testbed via JTAG, UART and Ethernet links from the host machine.
The GUI primarily relies on click-based configuration although the APIs could be wrapped into scripts to automate the testing.
The interface configuration function uses the UART and JTAG APIs to perform the configuration of a series of hardware registers, counters and special function blocks in the testbed and to perform serial status monitoring of different ECUs.
This link is also used to configure the Ethernet run-time debugger link between the host and the testbed.
The ECU configuration subsystem is used to load different applications and system configurations onto a target ECU, primarily through the JTAG APIs. 
These are invoked when loading specific \textit{elf} files to the ECU, to reset specific ECUs during a testing session and to periodically poll the status of the ECUs for updating on the GUI. 

\begin{table}[t]
\caption{Interfaces and APIs}
\begin{center}
\scalebox{0.9}{
\begin{tabular}{@{}p{2cm}p{1.8cm}p{5.0cm}@{}}
\toprule
\textbf{Link/Function} & \textbf{API} & \textbf{Description} \\
\midrule
UART/\$monitor & sys\_log() & Logs of basic system status, high-level errors, and IDS stats \\
UART/\$config & sys\_ctrl() & Transmits configuration parameters to setup ECU frequency, CAN bus mode, and default simulation settings \\
UART/\$attack & att\_ctrl() & Transmits specified injection attack content or delivers DoS attack and fuzzing attack commands to the corresponding attack core\\
UART/\$monitor & monitor() & low-resolution log of CAN bus and the IDS ECU signals \\
\midrule
JTAG/\$config & prog\_node()  & Download ELF to specific ECU, configure special nodes, and expose basic debugging \\
JTAG/\$control & reset\_node() & Apply reset sequence to the specific node or all nodes \\
JTAG/\$config & debug\_config() & Configure wave capture through bridge node. \\
\midrule
Ethernet/\$monitor & bus\_log() & Monitors CAN bus data at highest resolution. \\
Ethernet/\$monitor & signal\_capture() & Captures pre-defined signals such as virtual CAN bus levels, node data. \\
Ethernet/\$monitor & user\_capture() & Monitors user-defined byte or bit-level data. \\
\bottomrule
\end{tabular}}
\label{tab:interface_function}
\end{center}
\vspace{-6mm}
\end{table}

The interface configuration function is used to establish a reliable communication interface between the CAN testbed and the configuration/analysis features of the software system. 
It can be used to specify the interface parameters and to optionally set up high-speed connectivity between the host and the testbed. 
Once the connectivity is established, the ECU configuration subsystem is used to load different applications and system configurations onto a target ECU. 
The interface can be used to load specific \textit{elf} files to the ECU and/or to reset the ECU to a known good state if it encounters errors during the testing phase. 
Once configured and active, the status of the ECU and its parameters are periodically polled and updated on the GUI when a test is in progress. 

A key feature of our framework is the real-time monitoring of the test setup and internal signals from the hardware platform at very high granularity through a set of monitoring APIs. 
The monitoring APIs allow for reading specific status signals on demand or automatically reading them at a defined periodicity (default ~300\,ms) and logging them for further analysis. 
The API also allows simple operations to be applied to the status signals (such as conditional checks and logical operations) while being read from the hardware testbed to flag any anomalies during a testing session in near-real-time. 
Additionally, selected signals (from ECUs) and the bus activity are captured in real-time and transferred back to the host for visualising system conditions. 
The bridge node can be configured using the configuration APIs to capture a sequence of signals at the highest sampling rate (clock speed) and pack them as layer-2 Ethernet packets to be sent to the host. 
On the host, the receiving API will decode the signals from the packet, add them to a wave visualiser for real-time monitoring of the ECU states and the CAN network, and logs them for offline analysis, if specified. 
Finally, the attack injection and control are handled from the GUI which invokes dedicated APIs that communicate with the control node on the testbed. 
Using our framework, raw attack messages from openly available datasets can be replayed on the internal bus by the control node to recreate the attack scenario on our testbed. 
The APIs also allow for specific attack vector injections whereby the control node injects a targeted attack message on the CAN bus, to model spoofing attacks. 
Large-scale DoS and Fuzzing attacks can also be launched this way, although the dedicated node on the hardware can be used to inject them at higher speed by specifying a flag in the API call. 
The APIs can also load long sequences of attack vectors and benign messages (in a csv file, for instance) into the test environment from the GUI and trigger the logging system for automating long-form tests. 
An extension of the logging API can be additionally used to perform quick parsing of the captured CAN data bus to identify known attacks (such as flooding-based DoS), tag them using labels, and log them for further analysis. 
The tagging can be used to develop new attack datasets and/or to support/validate online training of supervised learning models with the hardware model-in-the-loop.

\subsection{Hardware Subsystem}
The hardware subsystem (testbed) of our framework implements the functional blocks of ECUs, the CAN bus, and specialised control blocks that implement attack injection, monitoring, and interfacing with the host PC.
For our deployment, we use AMD’s Artix-7 XC7A200T (Nexys Video board) as the FPGA platform to implement the testbed. 

\subsubsection{CAN Controller and Virtual CAN bus}
To model the CAN network, we modified the open-source CAN controller's host interface by adapting an AXI4 interface (for configuration by ECU) and exposing selected internal signals via a register interface to the bridge node.
At the bus end, we use buffers to connect \textit{tx} and \textit{rx} pins to the shared CAN bus (virtual bus) with tristate logic on the \textit{tx} lines to control transmission. 
The virtual bus implements the wired-AND logic behaviour of a physical CAN bus on twisted-pair connections, thus allowing all CAN protocol specifications such as inbuilt arbitration and error flagging on the virtual bus. 
The bus functions and CAN controller functionality were validated by wiring up to off-the-shelf CAN controllers. 
The virtual bus exposes the CAN signals via the Pmod interface on the Nexys board for external monitoring (via scope), wiring up to physical transceivers for bus expansion, and for scaling the testbed across multiple FPGA boards.

\subsubsection{ECU models}
In the testbed, ECUs are modeled around MicroBlaze soft processor and dedicated peripherals including our modified CAN controller. 
The main functions of the 4 ECU system we emulated for our tests are shown in Table~\ref{tab:ecu_descriptions}.
Each ECU receives an independent clock signal from a clock manager that can be configured to the required clock speed at startup (from the host PC).
At runtime, the software tasks on the ECU perform periodic (and event-driven) processing of sensor inputs and CAN messages and determines responses to be applied to actuators and/or to be sent via the CAN interface.
For instance, in Figure~\ref{attack} when \textit{ECU3} detects a collision from the sensor input or CAN message, \textit{ECU1} read a corresponding actuating message on the CAN bus to block the engine and transmission unit, and \textit{ECU2} trigger the airbags based on the CAN message to protect the passengers.
Additionally, an external reset can be applied to a specific ECU using the I/O switches on the Nexys platform. 
The software application for each ECU is implemented in bare-metal C and compiled using Vitis-SDK. 
Additionally, each ECU implements a task event counter as a life signal, the value of which is periodically transmitted as a CAN message to indicate its status. 
At the host, any deviation in the rate of increment can be detected as an additional load on the ECU during the testing phase.

\begin{table}[t]
\caption{Test ECU functions deployed for the evaluation}
\begin{center}
\begin{tabular}{@{}ll@{}}
\toprule
\textbf{ECU\#} & \textbf{Implemented Function} \\
\midrule
ECU1 & Engines and brake control unit \\
ECU2 & Airbag actuators and light sensor for auto headlights \\
ECU3 & Brake sensors and Collision detection sensors \\
ECU4 & Controls headlights, tail-lights and brake lights\\
\bottomrule
\end{tabular}
\label{tab:ecu_descriptions}
\end{center}
\end{table}

\subsubsection{FINN-IDS core}
To show the feasibility of validating IDS models using the framework, we integrated a 5-layer 4-bit quantised multi-layer perceptron (MLP) network described in~\cite{khandelwal2023exploring} as the IDS engine.
The model was trained and quantised using the brevitas quantisation aware training library~\cite{brevitas} and compiled to hardware using the FINN toolchain from AMD~\cite{umuroglu2017finn}.
The FINN flow generates an AXI-stream IP block of the model which is then integrated as a coupled accelerator to the ECU (case study-I) and also at the CAN controllers' host-interface logic (case study-II).
The specific MLP model was chosen as it provided state-of-the-art detection accuracy across multiple attack vectors with low resource- and energy-overhead for near-line-rate detection on CAN systems. 
Note that the same IDS IP block is used across both case studies to show the ability of the flexible platform to support different integration schemes and to perform trade-off evaluations in near-real-world conditions. 


\subsubsection{External integration \& extensions}
While our tests use self-contained ECUs with limited external I/O, the testbed can be extended to allow hardware sensors and actuators (or other I/O functions) to be connected to specific ECUs for a full-fledged HIL environment. 
Similarly, external standalone ECUs could be integrated into the test setup by connecting the virtual CAN bus to a CAN PHY and transceiver to support the line voltage on the physical layer. 
As mentioned before, the hardware platform can be extended through the pmod interface for multi-FPGA testbeds.
The software component can also scale in the above cases to enable real-time monitoring, logging and attack injection in the expanded test setup, and can be extended to support interfacing to multiple FPGAs independently to monitor all ECUs with relative ease.

\section{IDS integration case study}\label{casestudy}
\begin{figure*}[t]
  \centering  
  \includegraphics[width=1\textwidth]{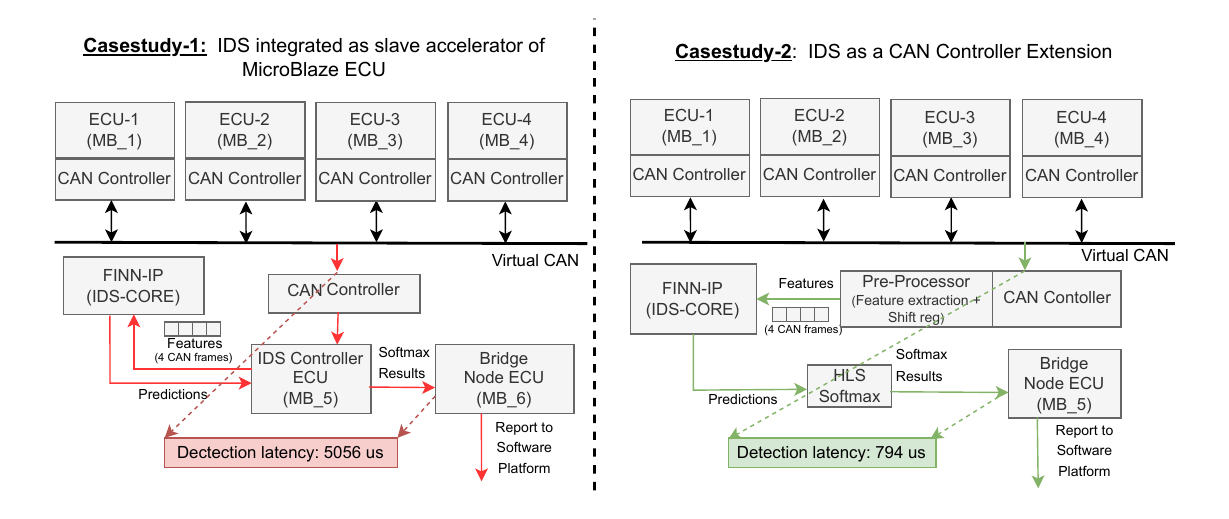}
  \caption{Overview of the CAN frames' datapath through the different IDS integrations (with \& without MicroBlaze ECU (MB) ) in the HIL environment. The left-hand side architecture presents a state-of-the-art coupled accelerator approach and the right-hand side architecture presents an integration approach as an extension of the CAN controller, with the coloured arrows indicating the flow of the CAN frame/features.}
  \label{smartcanids}
\end{figure*}
We used the Artix-7 XC7A200T on the Nexys Video development board as our testbed for the HIL simulator.
For our tests, the CAN controllers are configured for 500\, kbps operation, and the ECUs are clocked using independent 100\,MHz clocks.
WaveTrace was used to display the real-time waveform data received on the host machine.
We evaluate the capabilities by testing two IDS integration strategies progressing from an ECU-coupled accelerator to an embedded accelerator within the CAN controller's datapath.

\subsection{Case Study - I: IDS as a coupled accelerator}
In this case, the IDS model is integrated as a traditional memory-mapped accelerator to an ECU, MB\_5 in Fig.~\ref{smartcanids}, to form an IDS-enabled ECU. 
The IDS IP is attached as an AXI-streaming accelerator to the MicroBlaze processor, which is dedicated to executing the IDS task.
All other ECUs in the network perform a different function as described before. 
When a new message is available at the CAN interface of the IDS-enabled ECU, the message is read and pre-processed by a software task to generate a window of 4 messages, which is fed as the input feature to the IDS core to check for attack signatures. 
Once the IDS IP computes the result, a task on MB\_5 reads the result and applies a Softmax function on the 4 output values to arrive at the classification result. 
To display the result on the GUI, the 4-bit computed result is read by the \textit{bridge node} and passed to the software API. 


\subsection{Case Study - II: Extended CAN controller with IDS}
The coupled accelerator flow incurs software overheads and latency in moving the message from the CAN interface to ECU memory and subsequently to the accelerator and back, which could affect the performance of critical tasks on the ECU. 
We attempt to alleviate this latency and overhead by stitching the FINN core directly to the CAN controller's receive interface in case-II.
For simplicity,  we use a CAN receiver-only core, but the same approach could equally be applied with a full-fledged CAN controller. 
A hardware pre-processor logic parses the received messages directly off the host interface of the CAN receiver and uses a FIFO as a sliding window to accumulate 4 consecutive CAN messages, which is passed as the feature input to the IDS IP over the AXI streaming interface, stitching the IDS function at the network interface. 
Vitis-HLS is used to generate a hardware implementation of Softmax using DSP blocks. 
The resulting predictions are read by the \textit{bridge node} as before to report the observations to the software APIs. 

To compute the latency difference between the two approaches, we have instrumented the setup with timers which capture elapsed time from the start of a CAN frame on the bus to the completion of the Softmax function in both cases. 
Also, both cases use identical IDS IP to deploy our DNN-IDS function, with the final activation (Softmax) performed in software (in case 1) or through dedicated hardware (in case 2). 
This allows us to quantify the benefits and/or overheads of each integration approach in terms of IDS latency and hardware resource consumption.

\section{Results}
\label{results}

In this section, we present the different attack injection capabilities tested through the platform and observe the response of the ECU functions in the presence of these attacks. Subsequently, we integrate IDS functionality and evaluate the detection performance of the IDS model through the HIL framework.
We further quantify the detection latency of the IDS with both the integration strategies within testbed as discussed previously.
Finally, we quantify the resource overhead of the testbed and identify the overheads incurred for IDS integration in cases I and II respectively. 

\subsubsection{Attack simulation on ECUs}
To evaluate the attack injection capabilities, we simulate three different attacks, DoS, fuzzing, and spoofing, one each on three ECUs in the testbed. 
We initially observe the normal ECU behaviour under no attack conditions, and the expected responses from each ECU in the presence of control input(s) are recorded automatically through the setup. 
Subsequently, the attack conditions are triggered using the software interface, which injects the attack and automatically logs the network data and ECU responses for logging and analysis.
The details of the tests are presented in Table~\ref{tab:test_cases}.
Under all three attack conditions, normal ECU functionality cannot be carried out leading to critical functional loss such as no airbag deployment, braking action being terminated abruptly, and headlights toggling on/off without any user input.
All three scenarios can potentially threaten the safety of passengers under different conditions. 
Extended attacks were also injected through random injection of attack messages and looped testing through the software interface, where a similar loss of functionality was observed across the ECUs in the presence of attack messages. 
With the IDS integrated, the attack messages were positively flagged by the IDS for all three attack types, and we quantify the detection accuracy of the model using the framework next.

\begin{table}[t]
\caption{Expected v/s observed behaviours of ECUs under attack conditions.}
\begin{center}
\scalebox{0.9}{
\begin{tabular}{@{}p{1cm}p{1.5cm}p{2.0cm}p{4cm}@{}}
\toprule
\textbf{Test Case} & \textbf{Inputs} & \textbf{Expected Result} & \textbf{Under Attack} \\
\midrule
Collision signal & ECU3 activates collision detected signal & ECU2  activates airbags, ECU1 will disable engine control (EC) \& TCU & Under DoS attack, life signal lost, airbag cannot be activated, EC and TCU cannot be disabled, IDS reports threat\\
\midrule
Light control signal & ECU2 activates light sensor & ECU4 activates headlights and tail lights & Under fuzzing attack, lights activated without any light sensor input, IDS reports threat\\
\midrule
Brake signal & ECU3 senses brake signal & ECU1 will perform braking action ECU4 activates brake lights& Under spoofing attack, the braking action was unexpectedly terminated, Brake lights not activated,  IDS reports threat\\
\bottomrule
\end{tabular}}
\label{tab:test_cases}
\end{center}
\end{table}

\subsubsection{IDS Accuracy}
The QNN-IDS integrated for our test is a multi-class classification model that can detect DoS, fuzzing, and spoofing attacks along with benign messages.
For our tests, we replayed 180,000 test messages
from the openly available CAR Hacking dataset \cite{song2020vehicle}
which consisted of benign messages as well as different attack sequences, which were loaded as raw data for the test with automated monitoring and exporting enabled.
The model achieved a classification accuracy of 99.98\%, and the detailed confusion matrix of the classification result is shown in Table~\ref{table:confmatrix-4bit}.
Across the entire test set, the model only misclassified \textit{34} messages.
Furthermore, the model only misclassified 7 benign messages as false positives out of the total 103176 benign messages in the test set, which shows a low false alarm rate for the model.  
The detailed table showing accuracy metrics (\textit{Precision}, \textit{Recall} \& \textit{F1-score}) for all the attacks is shown in Table~\ref{table:accuracymetrics}.
The key capability here is that the model can be tested in an integrated environment and the validation can be automated through the framework to allow rapid prototyping of the IDS models in a real environment.

\begin{table}[t]
\centering
\caption{Confusion matrix capturing the classification results of the QNN-IDS.}
    \scalebox{1}{
        \begin{tabular}{llrrr}
            \toprule
            \multicolumn{1}{c}{} & \multicolumn{4}{c}{\textbf{Predicted Values}} \\
            \cmidrule{2-5}   
            \textbf{Attack Values} & \textbf{Benign}  & \textbf{DoS}  & \textbf{Fuzzing} & \textbf{RPM-Spoof}  \\ 
            \midrule
            \textbf{Benign}   & 103169     &        5                          &  2   &          0                      \\ 
            \textbf{DoS}      &   3  &     23690                             &     0    &        0                \\ 
            \textbf{Fuzzing}  &  23   &        0                        &      28065    &        1                  \\
            \textbf{RPM-Spoof}  & 0    &              0                 &        0  &   25042                     \\
            \bottomrule
        \end{tabular}}
\label{table:confmatrix-4bit}
\end{table}

\begin{table}[t]
\centering
\caption{Accuracy metrics of the QNN-IDS.}
\scalebox{1}{
\begin{tabular}{@{}lrrrr@{}}
\toprule
\textbf{Attacks}     & \textbf{Precicion} & \textbf{Recall} & \textbf{F1-Score} \\ \cmidrule{1-4}
DoS & 99.99 & 99.98 & 99.98 \\
Fuzzing & 99.99 & 99.91 & 99.95  \\
RPM-Spoof & 100 & 100  & 100  \\
\bottomrule
\end{tabular}}
\label{table:accuracymetrics}
\end{table}

\subsubsection{Latency \& Resource Utilisation}
We quantify the detection latency for both the IDS integration strategies discussed in the previous section.
For case study I, we observe the total detection time to be \textit{5,056 us}, from the message arriving at the CAN interface to the completion of the Softmax activation in the MB\_5 IDS-ECU.
For case study II, we observe more than 6$\times$ reduction in the detection latency with the classification result available in \textit{794 us} from the message arriving at the CAN interface to the completion of the hardware Softmax. 
The latency computation in both the case studies includes the message reading time from the CAN bus up to the final results reported back by the \textit{bridge node}.
Embedding the IDS closer to the controller alleviates software processing overheads (and delays) for preprocessing, data movement and Softmax activation, as shown in Fig.~\ref{smartcanids}. 
Offloading \textit{pre-processing} (ID, payload extraction $+$ message concatenation) and \textit{post-processing} (softmax) computations to hardware and integrating this close to the CAN controller removes the software bottleneck, and enables lower detection latency for line-rate IDS implementations.
For line rate detection on our 500\,kbits/s CAN network, we consider the acquisition window of 4 minimal-length CAN data frames ({296\,\textmu s each}) with protocol overheads.
In this case, the maximum latency for line rate detection should be {\textless\,1184\,\textmu s}.
From the tests using our HIL setup, we measure that IDS coupled to the CAN controller (case II) can perform detection at {794\,\textmu s} compared to the {5056\,\textmu s} incurred by the traditional ECU-coupled IDS scheme.
For higher speed CAN interfaces, the latency of the IDS IP can be further reduced through higher parallelization (unrolling) at the expense of higher energy and resource consumption.


We further quantify the hardware resource utilisation of all the common hardware components (ECUs, Control Node, Injection Node, Bridge Node, and the Debug Node) which are shared across both integration case studies in Table~\ref{tab:common_components}.
The resource consumption of the different IDS pathways is captured in Table~\ref{t3_6}.
The hardware offload consumes less general purpose and memory resources than the MicroBlaze IDS-ECU (IDS enabled ECU in the Table~\ref{t3_6}), while the hardware Softmax incurs DSP blocks to maximize performance. 
We observe a reduction of $\approx$1,200 LUTs (0.91 \%), $\approx$1,500 FFs (0.59 \%) and 16 BRAMs (4.3 \%) with the extended CAN controller scheme (CAN-coupled IDS in Table~\ref{t3_6}), compared to the coupled accelerator IDS-ECU method described in most literature. 
Additionally, the overall utilisation of the hardware subsystem (with 4 ECUs and IDS) is less than $\approx$65\% (BRAMs) of resources on the Artix-7 FPGA, making it possible to scale the hardware subsystem to incorporate more ECUs and faster data interfacing to the host,  with a single or multi-FPGA environment, making it an ideal solution for large-scale HIL validation and testing setup. 

\begin{table}[t]
\caption{Resource utilisation of hardware nodes used in the testbed.}
\begin{center}
\begin{tabular}{@{}lcccc@{}}
\toprule
\textbf{Functions} & \textbf{LUTs} & \textbf{FFs} & \textbf{BRAMs} & \textbf{DSPs} \\
\midrule
ECU1 & 3330 & 2664 & 32 & 0 \\
ECU2 & 3235 & 2653 & 32 & 0 \\
ECU3 & 3319 & 2758 & 32 & 0 \\
ECU4 & 3224 & 2614 & 32 & 0 \\
Control Node & 4549 & 3557 & 32 & 0 \\
DoS Node & 2658 & 2277 & 32 & 0 \\
Bridge & 2522 & 2713 & 32 & 0 \\
Debug & 736 & 1121 & 0.5 & 0 \\
Ethernet & 1079 & 1483 & 0 & 0 \\
\midrule
Total (\%)  & 24930 (18.5) & 23252 (8.6) & 224.5 (61.5) & 0 (0)\\
\bottomrule
\end{tabular}
\label{tab:common_components}
\end{center}
\end{table}

\begin{table}[t]
\caption{IDS resource utilisation for both case studies.}
\begin{center}
\scalebox{0.96}{%
\begin{tabular}{@{}lcccc@{}}
\toprule
\textbf{Function} & \textbf{LUTs} (\%) & \textbf{FFs} (\%) & \textbf{BRAMs} (\%) & \textbf{DSPs} (\%)\\
\midrule
\midrule
Case-I: IDS-ECU & 5990 (4.5) & 6297 (2.4) & 39 (10.6) & 0 (0) \\
Case-II: CAN-IDS & 4788 (3.7) & 4773 (1.8) & 23 (6.2) & 12 (1.6) \\
\bottomrule
\end{tabular}}
\label{t3_6}
\end{center} 
\end{table}

\section{Conclusion}\label{conclusion}
In this paper, we presented FAV-NSS, a hardware-in-loop test framework that can effectively emulate a multi-ECU setup with the CAN bus network protocol on an FPGA device. 
The software subsystem of FAV-NSS provides numerous capabilities for the user through graphical interfaces and APIs that enable controlling/automating test cases (injection of DoS, Fuzzing, Spoofing or other network attacks) on the CAN bus, real-time logging of test data (network, ECU status), and performing analysis and tagging of observed bus data at run-time.
The hardware subsystem models the ECUs using the MicroBlaze processor with a CAN controller as a memory-mapped peripheral. 
A virtual CAN bus connects together the CAN controllers and implements the wired-AND logic to model physical layer capabilities of the protocol.
To show the capabilities of the framework, we investigated multiple hardware integration strategies for an IDS solution, aiming to quantify the detection performance in terms of accuracy and latency. 
We used the software framework to automate the injection of attack vectors mixed with benign messages using a large test set of 180,000 CAN messages.
With the framework, the detection accuracy of the IDS in both integration strategies -- close to the network controller or as a coupled accelerator to an ECU -- was measured.
Additionally, the timing performance of both approaches was characterised by the instruments in the framework at runtime, with the network controller integration achieving 6.3$\times$ reduction in latency compared to traditional ECU-coupled IDS accelerator. 
In the future, we want to further enhance the communication method between software and hardware platforms to improve the capabilities of the framework specifically for multi-FPGA testbeds and to model ECUs using more complex processor architectures.

\bibliographystyle{IEEEtran}
\bibliography{references}

\end{document}